Non-uniform Matter in Neutron Star Crusts Studied by the Variational Method with Thomas-Fermi Calculations

H. Kanzawa[a], M. Takano [b,a], K. Oyamatsu[c, *] and K. Sumiyoshi[d]

a Department of Physics, Science and Engineering, Waseda University, 3-4-1 Okubo Shinjuku-ku, Tokyo 169-8555, Japan
b Research Institute for Science and Engineering, Waseda University, 3-4-1 Okubo Shinjuku-ku, Tokyo 169-8555, Japan
c Department of Media Theories and Production, Aichi Shukutoku University, Nagakute-cho, Aichi 480-1197, Japan
d Numazu College of Technology, Ooka 3600, Numazu, Shizuoka 410-8501, Japan

* Present address: Department of Library and Information Science, Aichi Shukutoku University, Nagakute-cho, Aichi 480-1197, Japan.

The equation of state (EOS) for neutron star (NS) crusts is studied in the Thomas-Fermi (TF) approximation using the EOS for uniform nuclear matter obtained by the variational method with the realistic nuclear Hamiltonian. The parameters associated with the nuclear three-body force, which are introduced to describe the saturation properties, are finely adjusted so that the TF calculations for isolated atomic nuclei reproduce the experimental data on masses and charge distributions satisfactorily. The resulting root-mean-square deviation of the masses from the experimental data for mass-measured nuclei is about 3 MeV. With use of the nuclear EOS thus determined, the nuclei in the crust of NS at zero temperature are calculated. The predicted proton numbers of the nuclei in the crust of NS are close to the gross behavior of the results by Negele and Vautherin, while they are larger than those for the EOS by Shen et al. due to the difference in the symmetry energy. The density profile of NS is calculated with the constructed EOS.

§ 1. Introduction

The nuclear equation of state (EOS) is an essential ingredient in the studies of astrophysical phenomena such as supernovae (SNe), hypernovae and black hole formations, and the influence of the nuclear EOS's on these phenomena has been studied.[1, 2, 3] At present, however, there are only two types of nuclear EOS's available for the simulations of these astrophysical phenomena. One is constructed by Lattimer and Swesty,[4] based on the compressible liquid drop model, and the other is constructed by Shen et al.[5] based on the relativistic mean field theory. Recently, the Shen-EOS is extended by Ishizuka et al. so as to treat hyperon mixing,[6] and also by Nakazato et al. to take into account quark-hadron phase transition.[2] It should be noted that these EOS's are constructed in phenomenological frameworks. Therefore, nuclear EOS's for SN simulations based on the many-body theories starting from the realistic nuclear Hamiltonian are desirable.

In this situation, we are now pursuing the construction of a nuclear EOS for SN simulations based on the realistic nuclear forces by the variational many-body calculation. As a first step in this project, we treated uniform nuclear matter in Ref. 7) (hereafter referred to as paper I): We calculated the (free) energies for symmetric nuclear matter and neutron matter at zero and finite temperatures using the cluster variational method with the Argonne v18 (AV18) two-body potential and Urbana IX (UIX) three-body potential. At zero temperature, the two-body energy without the three-body force was calculated in the two-body cluster approximation with the healing distance for the two-body correlation being chosen appropriately, and the contribution of the three-body force was taken into account in a simpler manner. The obtained energy per nucleon $E$ at zero temperature was in fair agreement with the results of the Fermi Hypernetted Chain (FHNC) variational calculation by Akmal, Pandharipande and Ravenhall (APR).[8]

At finite temperatures, we employed the method used by Schmidt and Pandharipande (SP).[9] In this method, the approximate orthogonality of the correlated basis (the Jastrow-type wave functions with different occupation probabilities of the single-particle states) is assumed. Recently, Mukherjee and Pandharipande investigated the validity of this method.[10]

The free energy obtained in paper I is close to the result by Friedman and Pandharipande (FP)[11] at low densities: At high densities, the free energy of paper I is appreciably higher than that by FP. The estimated critical temperature is about 18 MeV, which is close to the results of other studies.[12]

It should be noted that the formula to calculate $E$ includes adjustable parameters associated with the contribution from the nuclear three-body force (TBF) for the following reasons: (i) Compared with the well-determined modern two-body interaction, TBF has uncertainty. In fact, the UIX potential consists of the Fujita-Miyazawa-type two-$\pi$-exchange part $V_{ijk}^{2\pi}$ and the repulsive part $V_{ijk}^{R}$, the latter being rather phenomenological. According to the *ab initio* Green Function Monte Carlo calculations for light nuclei, no satisfactory reproduction of the experimental energy spectra is obtained with the UIX three-body potential, and then Pieper et al. proposed more sophisticated TBF models IL1~IL5.[13] The search for better TBF models is still in progress. (ii) In paper I, we evaluated the expectation value of the three-body Hamiltonian $H_3$ with the Fermi-gas wave function for simplicity. More sophisticated calculations of the TBF energy taking into account the correlation between nucleons as well as the possible relativistic correction will modify the expectation value of $H_3$. Considering these two points, we introduced, in paper I, adjustable parameters whose values are determined so as to reproduce the empirical saturation conditions.

Due to the above-mentioned treatment of the TBF energy, the variational method in paper I is more phenomenological than sophisticated microscopic many-body calculations such as the FHNC variational approach,[8] relativistic and nonrelativistic Brueckner theories[14] and quantum Monte Carlo calculations.[15] However, the simplicity of this variational method enables us to construct the SN-EOS, because the SN-EOS must cover an extremely wide range of mass densities $\rho$, temperatures $T$ and proton fractions $Y_p$ (In the case of the Shen-EOS, $10^{5.1}$ g/cm$^3$ $\leq \rho \leq 10^{15.4}$ g/cm$^3$, 0 MeV $\leq T \leq$ 100 MeV, $0 \leq Y_p \lesssim 0.56$).

In this paper, as the next step in our project, we construct the EOS for neutron star (NS) crusts, i.e., $\beta$-stable non-uniform nuclear matter at zero temperature. The NS crust is separated into the outer crust and the inner crust. In the outer crust, nucleons are confined in nuclei constructing the body-centered-cubic (BCC) lattice in a relativistic electron gas. In the inner crust, neutrons drip out of the nuclei. In this study, we treat the inhomogeneous matter appearing in the NS crust in the Thomas-Fermi (TF) approximation by following the route adopted by Shen et al.[5] or originally by Oyamatsu.[16]

Here, we pay attention to laboratory nuclei: Since experimental data are available on those nuclei, it is desirable that the constructed EOS reproduces those experimental data. Fortunately, in our theory, the adjustable parameters associated with the TBF contribution allow some freedom to readjust the EOS so that the TF calculation reproduces those experimental data. Therefore, as a preparation for the study of the NS crust, we perform the TF calculation for isolated atomic nuclei, and readjust the parameters in the EOS of uniform nuclear matter so that the TF calculation reproduces the experimental masses and radii of nuclei. Since the EOS constructed in paper I is reasonable, this readjustment must be rather small.

In §2, we redetermine the adjustable parameters in the EOS for uniform nuclear matter so as to reproduce the experimental data on the atomic nuclei. Using the parameter-readjusted energy per nucleon $E$, we construct the EOS in the neutron star crust in §3. A summary and conclusions are given in §4.

## § 2. Thomas-Fermi Calculations of Atomic Nuclei

*2.1. Thomas-Fermi approximation*

In this section, we perform a simplified version of the extended Thomas-Fermi (TF) calculations for atomic nuclei with the energy expression per nucleon $E$ for uniform nuclear matter constructed in paper I. In this process, we readjust the parameters in $E$ so as to reproduce the experimental data on atomic nuclei. We adopt the TF method that was originally proposed by Oyamatsu for the study of "pasta" nuclei in the inner crust of NS,[16] and also used by Shen et al. to construct the EOS for non-uniform SN matter.[5] Furthermore, it was recently adopted by Oyamatsu and Iida to investigate the relations between the gross feature of atomic nuclei and the EOS for uniform nuclear matter,[17] and also to study the "pasta" nuclei in the inner crust of NS in relation to the density dependence of the symmetry energy.[18]

In this TF approximation, the binding energy $B(Z, N)$ of a nucleus with the proton number $Z$ and the neutron number $N$ is given by

$$-B(Z, N) = \int d\boldsymbol{r}\, \varepsilon(n_p(r), n_n(r)) + F_0 \int d\boldsymbol{r} |\nabla n(r)|^2 + \frac{e^2}{2} \int d\boldsymbol{r} \int d\boldsymbol{r}' \frac{n_p(r) n_p(r')}{|\boldsymbol{r} - \boldsymbol{r}'|}. \quad (2.1)$$

The first term on the right-hand side of Eq. (2.1) is the bulk term. The energy density of the uniform nuclear matter $\varepsilon(n_p, n_n)$ with the proton number density $n_p$ and the neutron number density $n_n$ is given as the energy per nucleon $E$ multiplied by the total nucleon number density $n = n_p + n_n$. The second term on the right-hand side of Eq. (2.1) is the gradient term with $F_0$ being a parameter. The third term is the Coulomb energy.

The nucleus is assumed to be spherical, and the proton (neutron) number density $n_p(r)$ ($n_n(r)$) is assumed to be expressed as

$$n_i(r) = \begin{cases} n_i^{\text{in}} \left[1 - (r/R_i)^{t_i}\right]^3 & (0 \leq r \leq R_i), \\ 0 & (R_i \leq r). \end{cases} \quad (i = \text{p, n}) \quad (2.2)$$

Here, $n_i^{\text{in}}$ ($i = $ p or n for protons or neutrons, respectively) is the density at the center of the nucleus, $R_i$ is the "radius" of the nucleus, and $t_i$ is the parameter related to the surface diffuseness. There are six parameters with the following two normalization conditions:

$$Z = \int d\boldsymbol{r}\, n_\mathrm{p}(r), \quad N = \int d\boldsymbol{r}\, n_\mathrm{n}(r). \tag{2.3}$$

Therefore, $-B(Z, N)$ for each nucleus $(Z, N)$ is minimized with respect to four independent parameters in $n_i(r)$. Then, we obtain the mass of each nucleus, $M^{\mathrm{TF}}(Z, N)$, and the density distribution $n_\mathrm{p}(r)$, $n_\mathrm{n}(r)$.

*2.2. Many-body calculation for uniform nuclear matter*

The prescription given in paper I for constructing $E$ of uniform nuclear matter at zero temperature is as follows. First we decompose the nonrelativistic nuclear Hamiltonian $H$ into two parts: The two-body Hamiltonian $H_2$ including the isoscalar part of the AV18 two-body potential, and the three-body Hamiltonian $H_3$ composed of the UIX three-body potential.

The expectation value of $H_2$ is evaluated with the Jastrow-type wave function $\Psi$ in the two-body cluster approximation, and minimized with respect to the two-body correlation functions included in $\Psi$ to obtain the two-body energy per nucleon $E_2$. In this minimization, the healing-distance condition is imposed in addition to the extended Mayer's condition (normalization condition). A parameter $a_\mathrm{h}$ in the healing-distance condition is determined so that $E_2$ for symmetric nuclear matter is in good agreement with the result of FHNC calculation by APR.[8]

It should be noted that, in paper I, the nucleon mass $m_\mathrm{N}$ in the kinetic-energy operator in $H_2$ for symmetric nuclear matter was chosen as the average of the proton mass $m_\mathrm{p}$ and the neutron mass $m_\mathrm{n}$, i.e., $m_\mathrm{N} = (m_\mathrm{p} + m_\mathrm{n})/2$. In this paper, for simplicity, we fix it as $m_\mathrm{N} = m_\mathrm{n}$ for symmetric nuclear matter. As expected, we found that the shifts of the values of $E_2$ due to this replacement of $m_\mathrm{N}$ are negligibly small and that the corresponding wave function $\Psi$ as well as the parameter $a_\mathrm{h}$ remains virtually unchanged.

For the three-body energy, we decompose the UIX potential into the two-pion-exchange term $V_{ijk}^{2\pi}$ and the repulsive term $V_{ijk}^{R}$. Then, we express the three-body energy per nucleon $E_3$ as

$$E_3 = \alpha \frac{\langle H_3^R \rangle_F}{N_N} + \beta \frac{\langle H_3^{2\pi} \rangle_F}{N_N} + \gamma^2 \exp[-\delta n], \qquad (2.4)$$

where $N_N$ is the number of nucleons in nuclear matter, and

$$H_3^R = \sum_{i<j<k} V_{ijk}^R, \quad H_3^{2\pi} = \sum_{i<j<k} V_{ijk}^{2\pi}. \qquad (2.5)$$

The subscript F attached to the brackets in Eq. (2.4) represents the expectation value with the Fermi-gas wave function. The last term on the right-hand side of Eq. (2.4) is a correction term for symmetric nuclear matter ($\gamma = 0$ for neutron matter).

In Eq. (2.4), $\alpha$, $\beta$, $\gamma$ and $\delta$ are adjustable parameters. In paper I, we determined their values so that the obtained total energies per nucleon $E = E_2 + E_3$ for symmetric nuclear matter and neutron matter reproduce the empirical saturation density $n_0$, saturation energy $E_0$, incompressibility $K$ and symmetry energy $E_{sym}$.

*2.3. Energy for asymmetric nuclear matter*

For asymmetric nuclear matter, we calculate the energy per nucleon $E(n, x)$ for arbitrary proton fraction $x = n_p/n$ by interpolating between the energy for neutron matter $E(n, 0)$ and that for symmetric nuclear matter $E(n, 1/2)$. In this paper, we choose the following interpolation:

$$E(n,x) = E_F(n,x) + E_I\left(n, \frac{1}{2}\right) + \left[E_I(n,0) - E_I\left(n, \frac{1}{2}\right)\right](1-2x)^2. \qquad (2.6)$$

Here,

$$E_F(n,x) = \frac{3\hbar^2}{10 m_N}(3\pi^2 n)^{2/3}\left[x^{5/3} + (1-x)^{5/3}\right] \qquad (2.7)$$

is the Fermi-gas energy, and $E_I(n, x) = E(n, x) - E_F(n, x)$ for $x = 0$ and $1/2$. We note that the interpolation Eq. (2.6) is different from that used in paper I, in which we employed the following naïve interpolation:

$$E(n,x) = E\left(n, \frac{1}{2}\right) + \left[E(n,0) - E\left(n, \frac{1}{2}\right)\right](1-2x)^2. \qquad (2.8)$$

The dependence of $E(n, x)$ on the proton fraction has been studied in the variational methods.[19, 20, 21] Lagaris and Pandharipande showed in Ref. 19) that $E(n, x)$ based on the FHNC calculation is well approximated by the interpolation Eq. (2.6). Recently, we also calculated $E(n, x)$ for asymmetric nuclear matter by extending the variational method used in paper I, and found that the approximation Eq. (2.6) is better than Eq. (2.8).[21] We note that, in principle, the quadratic interpolation Eq. (2.6) is valid when a simplification $m_p = m_n$ is assumed as in Sec.2.2.

In this paper, we define the symmetry energy as the difference between the energy per nucleon for symmetric nuclear matter and for neutron matter, i.e., $E_{sym} = E(n, 0) - E(n, 1/2)$, at $n = n_0$, as in paper I. We note that, in the case of the interpolation Eq. (2.6), the symmetry energy defined above is different from that defined as follows:

$$E_{sym} = \frac{1}{8} \left.\frac{\partial^2 E(n,x)}{\partial x^2}\right|_{x=1/2}. \qquad (2.9)$$

For example, the symmetry energy for the EOS of paper I is $E_{sym} = 30$ MeV according to the present definition, while it is $E_{sym} = 29.2$ MeV by the definition Eq. (2.9).

*2.4.  Readjustment of the parameters in the three-body-potential energy*

The chief aim of this section is to redetermine the values of the parameters $\alpha$, $\beta$, $\gamma$ and $\delta$ in Eq. (2.4) in addition to $F_0$ in Eq. (2.1) so that the TF calculations reproduce the gross feature of the empirical nuclear data. In choosing the empirical data to use, we follow Refs. 16-18): We adopt the proton numbers $Z_\beta$, masses $M$ and root-mean-square (RMS) charge radii $R_c$ of nine nuclei on the smoothed $\beta$-stability line in the range of the mass numbers $25 \leq A \leq 245$ given in Table A.1 of Ref. 16). The theoretical RMS charge radius $R_c$ is obtained from the charge distribution of the nucleus $\rho_c(r)$, which is expressed, by taking into account the proton charge form factor assumed in the Gaussian form, as follows:

$$\rho_C(r) = \frac{1}{\left(\sqrt{\pi}a_p\right)^3} \int dr' \exp\left[-\frac{(r-r')^2}{a_p^2}\right] n_p(r'), \qquad (2.10)$$

with $a_p = 0.65$ fm.[22]

As mentioned in the last section, the gross feature of the EOS constructed in paper I is reasonable. Therefore, the readjustment of the parameters must not change this gross feature of the EOS too much: Actually, it changes the saturation energy only by a small amount.

*2.5. Numerical Results*

The values of $n_0$, $E_0$, $K$ and $E_{sym}$ after readjusting the parameters in the way mentioned above are shown in Table I. The corresponding values of the parameters $\alpha$, $\beta$, $\gamma$, $\delta$ as well as that of $F_0$ are shown in Table II. For comparison, the values before the readjustment are also shown in Tables I and II. The present value of the saturation energy $E_0$, which is mainly related to the masses of nuclei, is a little lower than that of paper I. The symmetry energy $E_{sym}$, which is mainly related to the proton and neutron drip lines as well as the $\beta$-stability line, increases only slightly, while the saturation density $n_0$, which is mainly related to the RMS charge radius, and the incomporessibility $K$ are the same as in paper I. The proton numbers $Z_\beta$, masses $M$ and RMS charge radii $R_c$ of nine nuclei on the smoothed $\beta$-stability line calculated with the parameter-readjusted EOS and the EOS of paper I are shown in Appendix A.

Figure 1 shows the energies per nucleon $E$ for uniform nuclear matter after the readjustment of the parameters (present EOS) and in paper I. For comparison, the energy per nucleon for uniform nuclear matter in the case of Shen-EOS is also shown. Since the difference between the present $E$ and that of paper I, which corresponds to the difference in the TBF energy $E_3$, is small, they are hardly distinguishable from each other in Fig. 1. On the other hand, the present $E$ for symmetric nuclear matter and neutron matter are lower than those of Shen-EOS except at low densities. Especially, the symmetry energy of Shen-EOS, $E_{sym} = 36.9$ MeV, is larger than that of the present EOS. Figure 2 shows the TBF energies $E_3$ of the present EOS and that of paper I. As expected, $E_3$ for symmetric nuclear matter is negative at low densities ($n \lesssim 0.18$ fm$^{-3}$) and positive at high densities so as to shift the saturation point of the two-body energy $E_2$ to the empirical one. The differences between $E_3$ of paper I and that of the present EOS are small both for neutron matter and symmetric nuclear matter. Actually, at $n = n_0$, the present $E_3$ is lowered only by about 0.15 MeV for neutron matter and by 0.24 MeV for symmetric nuclear matter. Therefore, the saturation energy $E_0$ is lowered a little and the symmetry energy $E_{sym}$ increases only slightly.

Although the lowering of $E_0$ is rather small, it brings about appreciable improvement in the results of the TF calculations. In Fig. 3 shown are $\Delta M = M^{TF} - M^{exp}$ for the mass-measured nuclei, where the superscript "TF" means the quantity obtained in the TF calculation and "exp" the experimental data.[23] For comparison, we also show the TF calculations with the EOS of Paper I with $F_0 = 68.65$ MeVfm$^5$, which is the value used by Oyamatsu for the EOS Model IV of Ref. 16). With the EOS of paper I, $\Delta M$ are positive for almost all the nuclei, and increase as the mass number $A$ increases (roughly, $\Delta M \sim 0.26A$ MeV), except for the shell staggering. Therefore, due to the rather small lowering of $E_0$ by 0.24 MeV, $|\Delta M|$ decreases significantly: The gross feature of the nuclear masses is reproduced by the TF calculation. The RMS deviation of the calculated masses from the experimental data with the present EOS is 3.09 MeV for 2219 nuclei with $Z$ and $N \geq 2$ (3.08 MeV for 2149 nuclei with $Z$ and $N \geq 8$), i.e., this deviation is comparable with those of the Weizsäcker-Bethe (WB) type mass formulas.

Figure 4 shows the proton and neutron drip lines as well as the $\beta$-stability lines with the present EOS and that of paper I. The calculated neutron and proton drip lines for the two cases are close to each other, and the $\beta$-stability lines for the two cases are almost indistinguishable from each other. The main reason is that the symmetry energy $E_{sym}$, which mainly governs the $\beta$-stability line and the drip lines, remains practically unchanged by the readjustment of the parameters as shown in Table I. Furthermore, since the value of $E_{sym}$ is appropriate as compared with the empirical value, the experimental $\beta$-stable nuclei are well reproduced, and the calculated neutron and proton drip lines are close to those for the sophisticated atomic mass formula constructed by Koura et al.[24]

Figure 5 shows the calculated charge distributions in $^{90}$Zr and $^{208}$Pb compared with the experimental data.[25] The obtained $\rho_c(r)$ with the present EOS and that of paper I are indistinguishable from each other, since the saturation density $n_0$ is unchanged by the parameter readjustment. It is also seen that the calculated $\rho_c(r)$ are in fairly good agreement with the experimental data.

## § 3. Crust of the neutron star

### 3.1. Thomas-Fermi calculations

In this section, we study the crust of NS using the parameter-readjusted EOS constructed in the last section. The NS crust is composed of protons, neutrons and electrons. In this paper, we treat the electrons as the uniform relativistic Fermi gas: Since the electron Fermi energy is much higher than the Coulomb energy, we neglect the screening effect caused by the non-uniform distribution of the electrons. The protons distribute non-uniformly to construct a lattice of proton clusters; the BCC lattice is preferable to minimize the Coulomb lattice energy.[16] When the average baryon number density $n_B$ is small, the neutrons are also located in and near the region where the protons exist. As $n_B$ increases, the neutron chemical potential $\mu_n$ increases. When $\mu_n$ becomes larger than the rest mass energy of the neutron $m_n c^2$, the neutrons drip out of the "nucleus". The density at which it first occurs is the critical density $n_c$, and the region of the crust with $n_B \leq n_c$ is the outer crust, while the region with $n_B \geq n_c$ is the inner crust.

In this paper, following the method by Shen et al.,[5] or originally by Oyamatsu[16], we treat the BCC lattice of the NS crust in the Wigner-Seitz (WS) approximation, assuming that the spherical nuclei are located at the centers of spherical WS cells. The average baryon density $n_B$ equals to $N_{cell}/V_{cell}$, where $N_{cell}$ is the nucleon number in the WS cell and $V_{cell}$ is the volume of the cell.

The energy in the WS cell, $E_{cell}$, is written as

$$E_{cell} = E_{nucl} + E_e + E_{Coul} + Z m_p c^2 + (N_{cell} - Z) m_n c^2. \qquad (3.1)$$

The first term on the right-hand side of Eq. (3.1) is the energy caused by the nuclear force

$$E_{nucl} = \int_{cell} dr\, \varepsilon(n_p(r), n_n(r)) + F_0 \int_{cell} dr\, |\nabla n(r)|^2, \qquad (3.2)$$

and $E_e$ is the energy of the relativistic electron gas:

$$E_e = \frac{m_e^4 c^5 a^3}{8\pi^2 \hbar^3} \left\{ x_e (2x_e^2 + 1)(x_e^2 + 1)^{1/2} - \ln\left[ x_e + (x_e^2 + 1)^{1/2} \right] \right\}, \quad (3.3)$$

with $m_e$ being the electron mass and

$$x_e = \frac{\hbar}{m_e c} (3\pi^2 n_e)^{1/3}. \quad (3.4)$$

In Eq. (3.3), $a$ is referred to as the lattice constant, which is defined by $a^3 = V_{\text{cell}}$. The third term on the right-hand side of Eq. (3.1) is the Coulomb energy in the WS cell, which is given by

$$E_{\text{Coul}} = \frac{e^2}{2} \int_{\text{cell}} d\mathbf{r} \int_{\text{cell}} d\mathbf{r}' \frac{[n_p(r) - n_e][n_p(r') - n_e]}{|\mathbf{r} - \mathbf{r}'|} + c_{\text{BCC}} \frac{(Ze)^2}{a}. \quad (3.5)$$

Here, the electron density $n_e$ is constant in the WS cell as assumed above. The second term on the right-hand side of Eq. (3.5) is the correction term for the BCC lattice with $c_{\text{BCC}} = 0.006562$[16] and with $Z$ being the proton number per cell. We note that $Z = n_e V_{\text{cell}}$ because of the charge neutrality.

In the calculation of $E_{\text{cell}}$, the density distribution of the nucleons in the WS cell is assumed as follows:

$$n_i(r) = \begin{cases} (n_i^{\text{in}} - n_i^{\text{out}})\left[1 - (r/R_i)^{t_i}\right]^3 + n_i^{\text{out}} & (0 \leq r \leq R_i), \\ n_i^{\text{out}} & (R_i \leq r \leq R_{\text{cell}}), \end{cases} \quad (i = \text{p}, \text{n}) \quad (3.6)$$

with $R_{\text{cell}}$ being the radius of the WS cell: $R_{\text{cell}} \equiv (3V_{\text{cell}}/4\pi)^{1/3}$. It is noted that $n_p^{\text{out}} = 0$ both for the outer crust and the inner crust, while $n_n^{\text{out}} = 0$ only in the outer crust: By definition, $n_n^{\text{out}} > 0$ in the inner crust. Then, for a given $n_B$, the average energy density $E_{\text{cell}}/V_{\text{cell}}$ is minimized with respect to the parameters in the expressions of $n_p(r)$ and $n_n(r)$ in Eq. (3.6). It should be emphasized that the energy expression Eq. (3.1) and the parameterization of $n_i(r)$ in Eq. (3.6) are consistent with those adopted in the TF calculation of atomic nuclei in the last section.

*3.2. Numerical results*

   In this subsection, we present our numerical results in comparison with those by Baym, Pethick and Sutherland (BPS)[26] and by Negele and Vautherin (NV),[27] as well as the results with Shen-EOS. The EOS by BPS constructed with use of a mass formula and that by NV calculated in the Hartree-Fock approximation are typical EOS's for the outer crust and the inner crust, respectively; both have been used for NS studies by many researchers. In addition, Shen-EOS constructed recently has also been used in many astrophysical studies as a typical SN-EOS.

   The nuclei appearing in the outer crust of NS with the present (parameter-readjusted) EOS are shown in Table III. Also shown are the results by BPS and with Shen-EOS. The result with the present EOS is close to the gross behavior of BPS, though there is no preference of the magic nuclei with $Z = 28$, $N = 50$ and $N = 82$ in our result, in contrast with the result by BPS which includes the shell correction. In the case of Shen-EOS, the neutron drip occurs at smaller $Z$. The main reason is the larger $E_{sym}$ of Shen-EOS as will be discussed below. The critical density at which $\mu_n = m_n c^2$ is evaluated for the present EOS as $n_c = 2.61 \times 10^{-4}$ fm$^{-3}$.

   Figure 6 shows the proton number $Z$ in the inner crust as a function of $n_B$. For comparison, we also show the results with the EOS of paper I, Shen-EOS and those by NV. The present result is close to the gross behavior of the result by NV, though the clear magic and sub-magic numbers $Z = 50$ and 40 due to the shell effect appearing in the result by NV are not predicted in the present calculation.

   In Fig. 6, a significant difference is seen between the result with the present EOS and that with Shen-EOS. According to the recent systematic study of the nuclei in the inner crust with the TF calculation by Oyamatsu and Iida,[18] the proton number $Z$ in the inner crust decreases as the symmetry energy $E_{sym}$ of the EOS $\varepsilon(n_p, n_n)$ used in Eq. (3.2) increases, or, more precisely, as the symmetry energy density derivative coefficient $L = 3n[dE_{sym}(n)/dn]$ at $n = n_0$ of the EOS increases. In the case of the Shen-EOS, $E_{sym} = 36.9$ MeV and $L = 49.5$ MeV, which are larger than $E_{sym} = 30.0$ MeV and $L = 33.9$ MeV for the present EOS. Therefore, the result shown in Fig. 6 is consistent with their conclusion.

   The calculated $Z$ with the EOS of paper I is not very different from that with the present EOS, because $E_\text{sym}$ of these two EOS's are very close to each other, as shown in Table I: More precisely, $L$ for the present EOS is close to $L = 34.5$ MeV for the EOS of paper I. It should be noted here that the difference between the result with the present EOS and that of paper I is attributable to the uncertainty in the TBF energy: The influence of the uncertainty in the TBF energy is rather small on the TF calculation for the NS crust.

According to Ref. 18), the "pasta" nuclei appear when $L \lesssim 100$ MeV. Since $L = 33.9$ MeV in the case of the present EOS, the "pasta" phase is expected between the spherical-nucleus phase in the inner crust and the core of NS. However, we neglect the possible "pasta" phase, for simplicity, in this paper: The treatment of the "pasta" phase is the future problem.

Figure 7 shows the proton fraction averaged over the WS cell in the non-uniform region, $Y_\text{p}$, as a function of $n_\text{B}$. For comparison, the same quantities with the EOS of paper I and Shen-EOS are also shown. The two vertical boundary lines in this figure are pertinent to the present EOS. In the core region, possible muon mixing is taken into account, where the muons are treated as the relativistic Fermi gas. Actually, muon mixing occurs at the densities higher than 0.137 fm$^{-3}$ for the present EOS. In addition, the critical density above which the direct URCA process occurs ($Y_\text{p} \geq 1/9$) is about 1.0 fm$^{-3}$ for the present EOS. It is noted that, the results with the present EOS and that of paper I are hardly distinguishable from each other: The difference in $Y_\text{p}$ is less than 0.02. This implies that the influence of the uncertainty in the TBF energy on the proton fraction in NS is small. On the other hand, the result with the present EOS in the inner crust is larger than that with Shen-EOS and, in the core region, $Y_\text{p}$ with the present EOS is much smaller than that with Shen-EOS. This feature is also related to the values of $L$ for the two EOS's; $L$ of the present EOS is considerably smaller than that of Shen-EOS. Since $L$ corresponds to the derivative of the symmetry energy $E_\text{sym}(n)$ with respect to density, $E_\text{sym}(n)$ for the present EOS is larger than that for Shen-EOS at subnuclear densities ($n \lesssim 0.1$ fm$^{-3}$, see Fig. 1), making the present $Y_\text{p}$ at these densities larger than that with Shen-EOS. As the density increases, the present $E_\text{sym}(n)$ becomes smaller than that for Shen-EOS and correspondingly the present $Y_\text{p}$ becomes smaller than that with Shen-EOS.

Finally, we solve the Tolman-Oppenheimer-Volkoff equation with the present nuclear EOS to obtain the neutron star profiles. The mass of the neutron star calculated with the present EOS is shown in Fig. 8. The maximum mass of the neutron star is 2.20 $M_\odot$, and the causality is violated at the central matter density $\rho_{m0}$ higher than $\rho_{mc} = 1.91 \times 10^{15}$ g/cm$^3$, at which the mass of the star is 2.17 $M_\odot$. Figure 9 shows the density profile of the neutron star having a mass of 1.4 $M_\odot$ with the present EOS. The boundary between the core and the inner crust, and the neutron drip point are also shown with the vertical lines in this figure. The obtained density profile of the neutron star is quite reasonable.

§ 4. Summary and conclusions

In this paper, we constructed the EOS for the crust and core of the neutron star based on the EOS for uniform nuclear matter that was obtained in paper I with the realistic two-body and three-body nuclear potentials. Making use of the uncertainty of the TBF energy $E_3$, we readjusted the parameters in $E_3$ a little so that the TF calculation for isolated atomic nuclei reproduces the gross feature of the experimental data on their masses and charge distributions. After readjusting the parameters in $E_3$, the RMS deviation of the nuclear masses from the experimental data is 3.09 MeV for 2219 nuclei with $Z$ and $N \geq 2$, which is comparable to that of the WB mass formula. The calculated $\beta$-stability line is in good agreement with the locations of the experimental $\beta$-stable nuclei, and the proton and neutron drip lines in the present calculations are close to those with the sophisticated mass formula by Koura et al. The charge distributions in $^{90}$Zr and $^{208}$Pb are in fairly good agreement with the experimental data. With use of this parameter-readjusted EOS, the nuclei in the outer crust of NS are calculated in the TF approximation, and compared with the results by BPS and Shen et al. In the inner crust, the proton number of the "nucleus" is compared with those by NV and Shen-EOS. The present result is close to the gross behavior of the result by NV, while it is larger than that by Shen-EOS due to the difference of the symmetry energy, or the density symmetry coefficient $L$, as has been pointed out by Oyamatsu and Iida.[18] On the other hand, the difference between the result for the interior properties of NS with the present EOS and that with the EOS of paper I, which is attributable to the uncertainty in the TBF energy, is rather small. Finally, the mass and the density profile of NS are calculated with the present EOS.

The obtained nuclear EOS is unique in the following point of view. This EOS is directly connected with the bare nuclear forces AV18 and UIX through our variational method, being different from other phenomenological approaches for the SN-EOS. Furthermore, in this study, the contribution from TBF, which has some uncertainty, is fixed so as to reproduce the gross feature of the atomic nuclei in the TF approximation, providing a more reliable EOS at subnuclear densities. We emphasize that this feature of the present EOS is especially important in our future work in constructing the nuclear EOS for SN simulations, in which the thermodynamic quantities as well as the chemical compositions of the matter from very low density ($\sim 10^5$ g/cm$^3$) to very high density ($\sim 10^{15}$ g/cm$^3$) are necessary, as mentioned in §1. The present EOS constructed in this paper, which provides reasonable description of low density matter, is more suitable for our project than the EOS of paper I. In this paper, it is shown that the present EOS for the crust of NS is close to the gross behavior of the results by BPS and NV. In order to discuss the compositions of the NS crusts in more detail, it is necessary to take into account quantum corrections such as shell and pairing effects; this is an interesting future problem.

As mentioned in §1, the variational method adopted in this study is relatively simple including some phenomenological treatment. To use more refined and less phenomenological variational method is an important future problem; an attempt to refine the calculations of the TBF energy by taking into account correlations between nucleons was reported in Ref. 28).

Toward the SN-EOS, a similar TF calculation is necessary in the wide range of density, temperature and proton fraction. As in the case of the Shen-EOS, free energies at finite temperatures should be used in the TF calculation. Since the free energies of uniform nuclear matter at finite temperature are constructed in paper I, at least in principle, we have reached the stage to proceed to the TF calculation of hot, inhomogeneous SN matter, though it will be a laborious task. In that calculation, we need the free energy of uniform asymmetric nuclear matter, for which the variational calculation will be reported elsewhere.

Acknowledgments

We would like to express special thanks to Profs. S. Yamada, M. Yamada, K. Iida, H. Nakada, A. Ohnishi, Y. Akaishi and H. Suzuki for valuable discussions. This study is supported by a Grant-in-Aid for the 21st century COE program "Holistic Research and Education Center for Physics of Self-organizing Systems" at Waseda University, and Grants-in-Aid from the Scientific Research Fund of the JSPS (Nos. 18540291, 18540295, 19540252, 21540280) and the Grant-in-Aid for Scientific Research on Innovative Areas (No. 20105004).

Appendix A

------- Readjustment of the parameters in the three-body energy $E_3$ --------

We readjust the parameters $\alpha$, $\beta$, $\gamma$ and $\delta$ in the three-body energy $E_3$ shown in Eq. (2.4) so that the TF calculations for proton numbers $Z_\beta$, masses $M$ and RMS charge radii $R_c$ of nine $\beta$-stable nuclei better reproduce the smoothed empirical data shown in Table A.1 of Ref. 16). The results are shown in Table A.1. Also shown are the results with the EOS of paper I with $F_0 = 68.65$ MeVfm$^5$ taken from Ref. 16). As seen in this Table, a satisfactory reproduction is obtained by this readjustment with $|\Delta Z_\beta| \lesssim 1$, $|\Delta Mc^2| < 2$ MeV and $|\Delta R_c| < 0.01$ fm.

References
1) K. Sumiyoshi, S. Yamada and H. Suzuki, Astrophys. J. 667 (2007) 382.
2) K. Nakazato, K. Sumiyoshi and S. Yamada, Phys. Rev. D 77 (2008) 103006.
3) K. Sumiyoshi, C. Ishizuka, A. Ohnishi, S. Yamada and H. Suzuki, Astrophys. J. 690 (2009) L43.
4) J. M. Lattimer and F. D. Swesty, Nucl. Phys. A 535 (1991) 331.
5) H. Shen, H. Toki, K. Oyamatsu and K. Sumiyoshi, Nucl. Phys. A 637 (1998) 435.
H. Shen, H. Toki, K. Oyamatsu and K. Sumiyoshi, Prog. Theor. Phys. 100 (1998) 1013.
6) C. Ishizuka, A. Ohnishi, K. Tsubakihara, K. Sumiyoshi and S. Yamada, J. Phys. G 35 (2008) 085201.
7) H. Kanzawa, K. Oyamatsu, K. Sumiyoshi and M. Takano, Nucl. Phys. A 791 (2007) 232.
8) A. Akmal, V. R. Pandharipande and D. G. Ravenhall, Phys. Rev. C 58 (1998) 1804.
9) K. E. Schmidt and V. R. Pandharipande, Phys. Lett. 87 B (1979) 11.
10) A. Mukherjee and V. R. Pandharipande Phys. Rev. C75 (2007) 035802.
11) B. Friedman and V. R. Pandharipande, Nucl. Phys. A 361 (1981) 502.
12) A. Rios, A. Polls, A. Ramos and H. Muther, Phys. Rev. C 78 (2008) 044314, and references therein.
13) S. C. Pieper, V. R. Pandharipande and R. B. Wiringa, Phys. Rev. C 64 (2001) 014001.
14) M. Baldo and C. Maieron, J. Phys. G 34 (2007) R243, and references therein.
15) S. Gandolfi, F. Pederiva, S. Fantoni and K. E. Schmidt, Phys. Rev. Lett. 98 (2007) 102503.
A. Sarsa, S. Fantoni, K. E. Schmidt and F. Pederiva, Phys. Rev. C68 (2003) 024308.
J. Carlson, J. Morales Jr., V. R. Pandharipande and D. G. Ravenhall, Phys. Rev. C68 (2003) 025802.
16) K. Oyamatsu, Nucl. Phys. A 561 (1993) 431.
17) K. Oyamatsu and K. Iida, Prog. Theor. Phys. 109 (2003) 631.
18) K. Oyamatsu and K. Iida, Phys. Rev. C 75 (2005) 015801.
19) I. E. Lagaris and V. R. Pandharipande, Nucl. Phys. A 369 (1981) 470.

20) M. Takano and M. Yamada, Prog. Theor. Phys. 116 (2006) 545.
21) M. Takano et al., in preparation.
22) L. R. B. Elton and A. Swift, Nucl. Phys. A94 (1967) 52.
23) G. Audi, A. H. Wapstra and C. Thibault, Nucl. Phys. A 729 (2003) 337.
24) H. Koura, T. Tachibana, M. Uno and M. Yamada, Prog. Theor. Phys. 113 (2005) 305.
H. Koura, M. Uno, T. Tachibana and M. Yamada, Nucl. Phys A 674 (2000), 47.
25) H. de Vries, C. W. de Jager and C. de Vries, At. Data nucl. Data Tables 36 (1987) 495.
26) G. Baym, C. J. Pethick and P. Sutherland, Astrophys. J. 170 (1971) 299.
27) J. W. Negele and D. Vautherin, Nucl. Phys. A207 (1973) 298.
28) H. Kanzawa, K. Oyamatsu, K. Sumiyoshi and M. Takano, AIP Conf. Proc. 1011 (2008) 287.

Figure captions

Fig. 1: Energies per nucleon $E$ for symmetric nuclear matter and neutron matter as functions of the nucleon number density $n$. The parameter-readjusted $E$ (Present) and that of paper I are hardly distinguishable from each other. For comparison, $E$ in the case of Shen-EOS are also shown.

Fig. 2: The three-body potential energies $E_3$ for symmetric nuclear matter and neutron matter as functions of the nucleon number density $n$.

Fig. 3: Deviation of the masses of the mass-measured nuclei $\Delta M = M^{TF} - M^{exp}$, where $M^{TF}$ is the mass by the TF calculation and $M^{exp}$ is the experimental data. The upper panel shows the results with the EOS of paper I, while the lower panel is in the case of the present EOS.

Fig. 4: The proton and neutron drip lines and the $\beta$-stability lines in the TF calculations with the EOS of paper I and the present EOS. Also shown are the experimental $\beta$-stable nuclei. The crosses represent the neutron and proton drip lines (with even-odd staggering) calculated with the mass formula by Koura et al.[24]

Fig. 5: Charge distributions in $^{90}$Zr and $^{208}$Pb calculated with the present EOS and the EOS of paper I, compared with the experimental data. The calculated charge distributions with these EOS's are hardly distinguishable from each other.

Fig. 6: The proton number in the inner crust of NS. Also shown are the results with the EOS of paper I, Shen-EOS and by NV.

Fig. 7: The proton fraction in NS as a function of the baryon density. Also shown are the results with the EOS of paper I and Shen-EOS.

Fig. 8: The mass of the neutron star as a function of the central matter density $\rho_{m0}$ with the present EOS. The vertical dashed line shows the critical density above which the causality is violated.

Fig. 9: The density profile of the neutron star having a mass of 1.4 $M_\odot$ with the present EOS.

Table I: The saturation densities $n_0$, saturation energies $E_0$, incompressibilities $K$ and symmetry energies $E_{sym}$ of uniform nuclear matter for the EOS of paper I and the present EOS. The numbers in the parentheses are the symmetry energies calculated with Eq. (2.9).

Table II: The parameters $\alpha$, $\beta$, $\gamma$ and $\delta$ in the three-body energy $E_3$ for the EOS of paper I and the present EOS. Also given is the coefficient of the gradient term, $F_0$.

Table III: The nuclei in the outer crust of the neutron star. The results in the TF calculations with the present EOS and Shen-EOS are shown in addition to those by BPS. The critical density $n_c$ is $2.61 \times 10^{-4}$ fm$^{-3}$ for the present EOS: In the case of Shen-EOS, $n_c = 1.90 \times 10^{-4}$ fm$^{-3}$.

Table A. 1: The empirical proton numbers, mass excesses and RMS charge radii on the smoothed $\beta$-stability line. Following the mass numbers, listed are the input data taken from Table A.1 of Ref. 16), and the results with the EOS of paper I and the present EOS.

Table I

|  | $n_0$ [fm$^{-3}$] | $E_0$ [MeV] | $K$ [MeV] | $E_{sym}$ [MeV] |
|---|---|---|---|---|
| **Paper I** | 0.16 | −15.85 | 250 | 29.91  (29.21) |
| **Present** | 0.16 | −16.09 | 250 | 30.00  (29.29) |

Table II

|  | $\alpha$ | $\beta$ | $\gamma$ [MeV·fm$^6$] | $\delta$ [fm$^3$] | $F_0$ [MeV·fm$^5$] |
|---|---|---|---|---|---|
| **Paper I** | 0.41 | −0.22 | −1604 | 13.93 | --------- |
| **Present** | 0.43 | −0.34 | −1804 | 14.62 | 68.00 |

Table III

| $n_b$ [fm$^{-3}$] | Present | | Shen-EOS | | BPS | |
|---|---|---|---|---|---|---|
| | N | Z | N | Z | N | Z |
| 9.98E−11 | 35.88 | 28.89 | 37.06 | 28.99 | 30 | 26 |
| 1.58E−9 | 36.18 | 29.00 | 37.29 | 29.12 | 30 | 26 |
| 1.99E−8 | 37.10 | 29.29 | 38.08 | 29.31 | 34 | 28 |
| 1.58E−7 | 38.89 | 29.83 | 39.70 | 29.70 | 34 | 28 |
| 3.97E−6 | 46.44 | 31.88 | 46.30 | 31.24 | 50 | 34 |
| 1.26E−5 | 52.27 | 33.26 | 51.38 | 32.14 | 50 | 32 |
| 7.92E−5 | 69.61 | 36.61 | 65.39 | 33.76 | 50 | 28 |
| 1.10E−4 | 74.41 | 37.37 | 69.43 | 34.06 | 82 | 42 |
| 2.57E−4 | 90.67 | 39.47 | | | 82 | 36 |
| 2.61E−4 | 91.02 | 39.51 | | | | |

Table A.1

|   | input | | |
|---|---|---|---|
| A | data | Paper I | Present |
| (a) proton number | | | |
| 25 | 12.41 | 11.75 | 11.76 |
| 47 | 21.86 | 21.40 | 21.42 |
| 71 | 31.70 | 31.43 | 31.46 |
| 105 | 45.09 | 44.96 | 45.02 |
| 137 | 57.14 | 57.11 | 57.20 |
| 169 | 68.85 | 68.78 | 68.89 |
| 199 | 79.61 | 79.33 | 79.48 |
| 225 | 89.18 | 88.22 | 88.38 |
| 245 | 96.39 | 94.90 | 95.08 |
| (b) mass excess in MeV | | | |
| 25 | −13.10 | −4.69 | −11.15 |
| 47 | −46.17 | −32.71 | −44.85 |
| 71 | −72.38 | −53.89 | −72.13 |
| 105 | −89.69 | −63.27 | −90.06 |
| 137 | −84.89 | −50.71 | −85.45 |
| 169 | −61.18 | −19.07 | −61.69 |
| 199 | −23.12 | 26.42 | −23.53 |
| 225 | 21.22 | 77.27 | 21.02 |
| 245 | 61.21 | 123.13 | 62.06 |
| (c) root-mean-square radius in fm | | | |
| 25 | 3.029 | 3.026 | 3.022 |
| 47 | 3.567 | 3.566 | 3.565 |
| 71 | 3.997 | 4.003 | 4.003 |
| 105 | 4.487 | 4.489 | 4.491 |
| 137 | 4.874 | 4.867 | 4.870 |
| 169 | 5.206 | 5.194 | 5.198 |
| 199 | 5.466 | 5.469 | 5.474 |

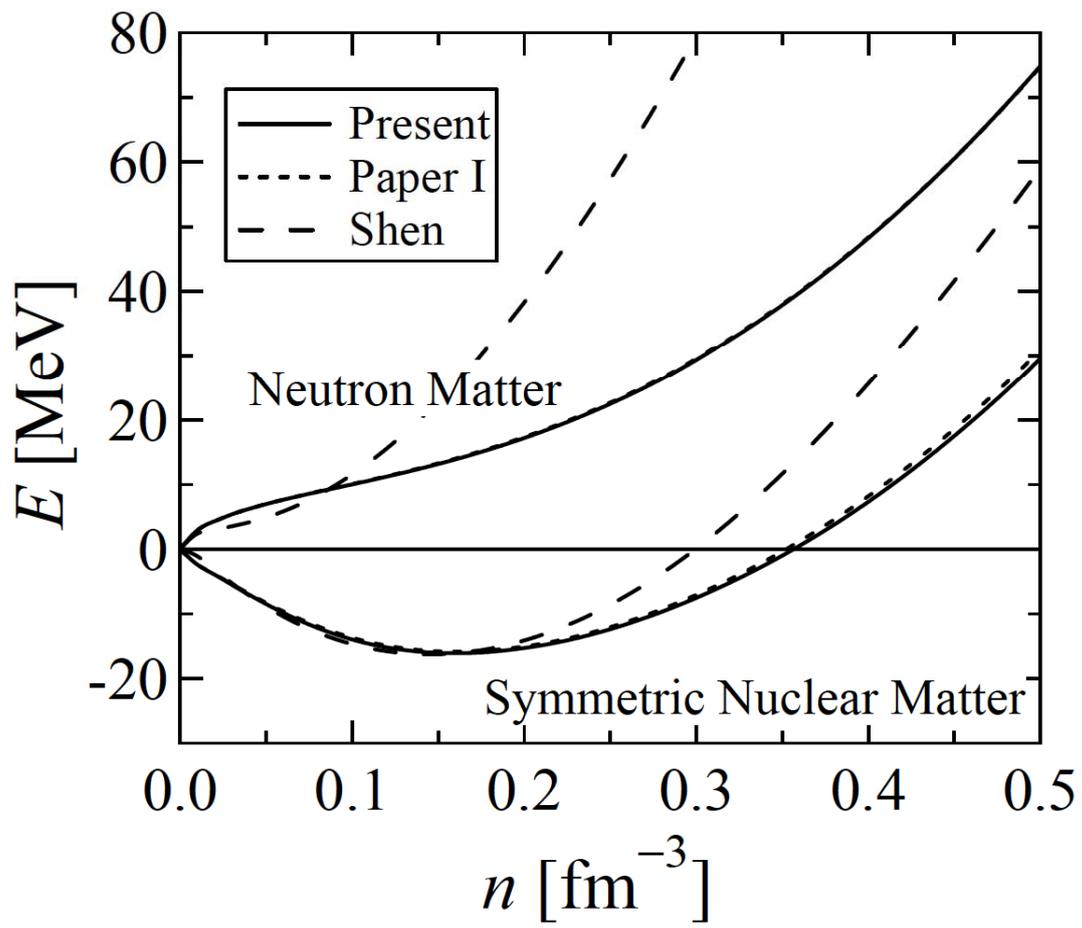

Figure 1

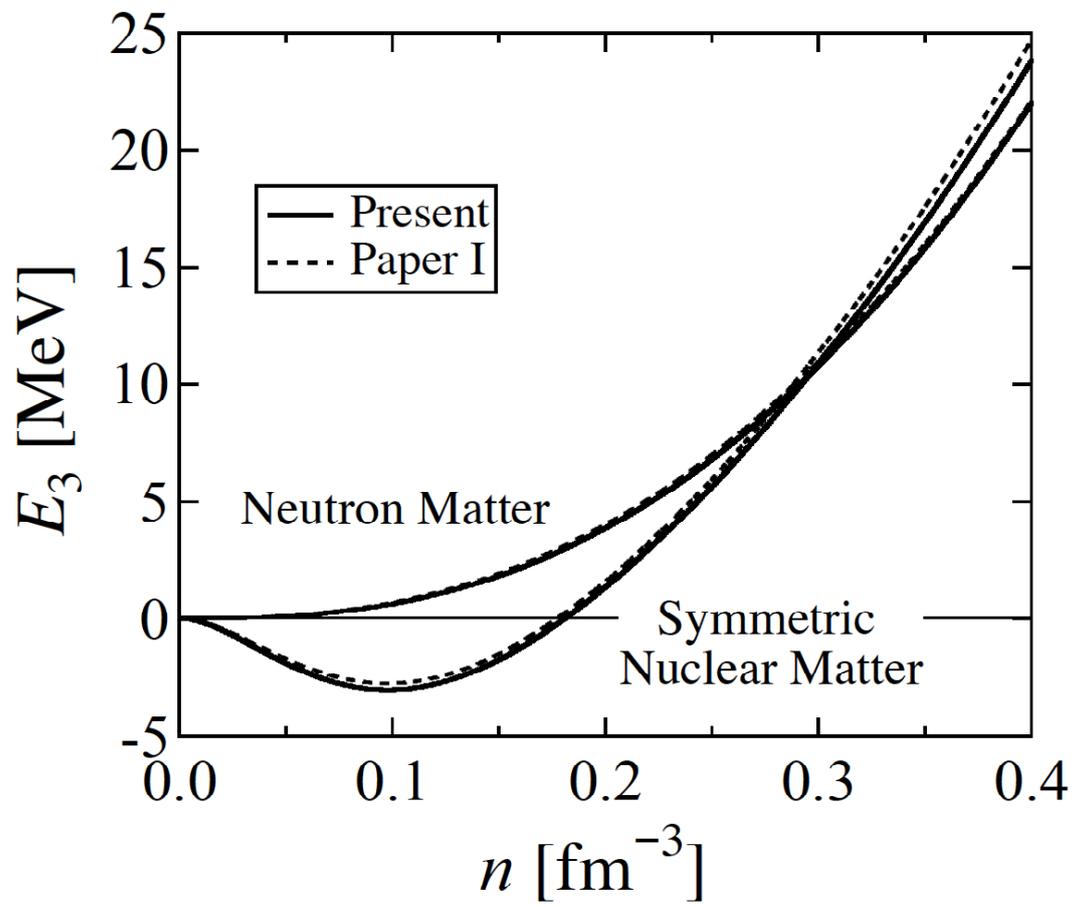

Fig. 2

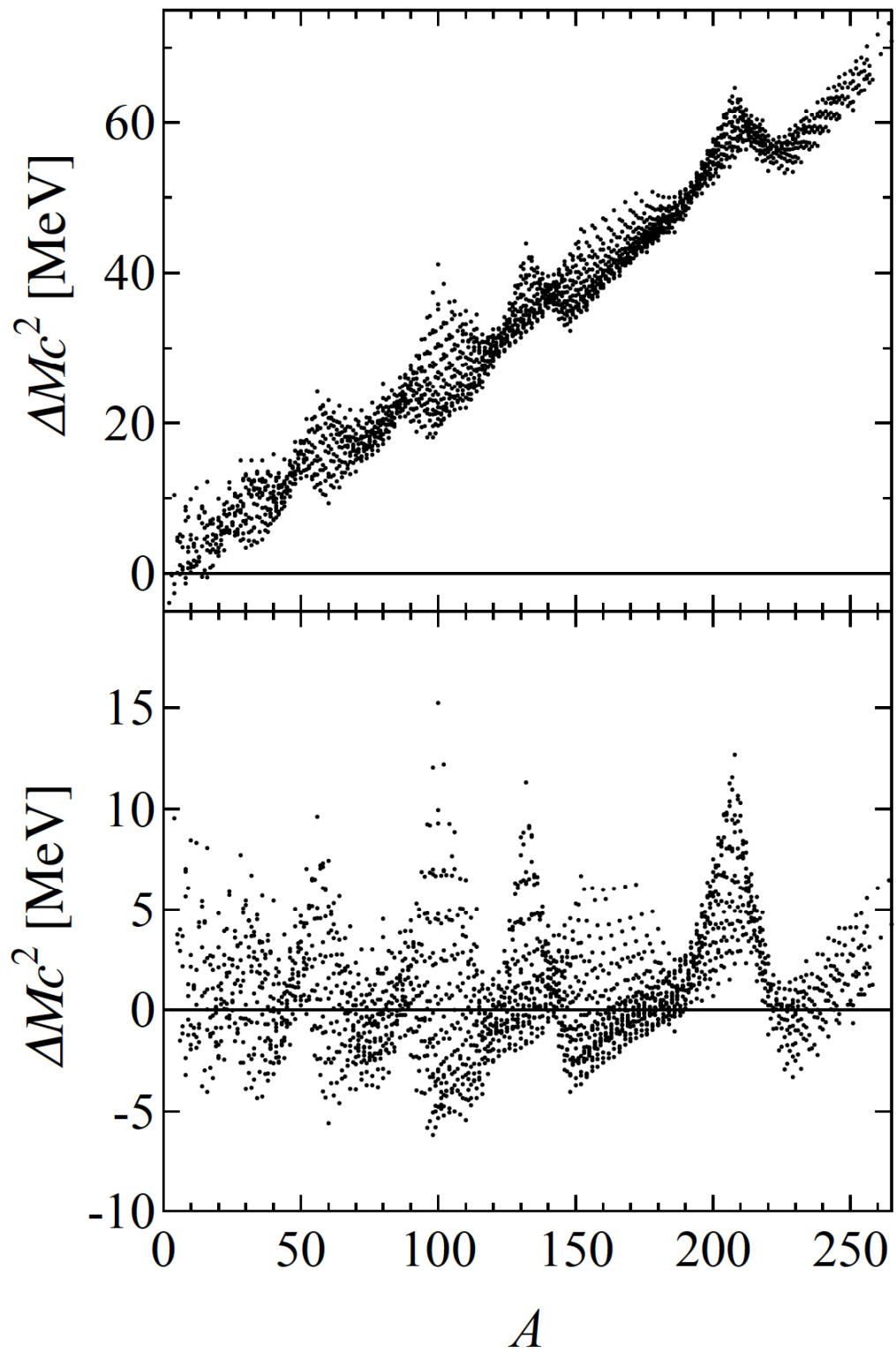

Fig. 3

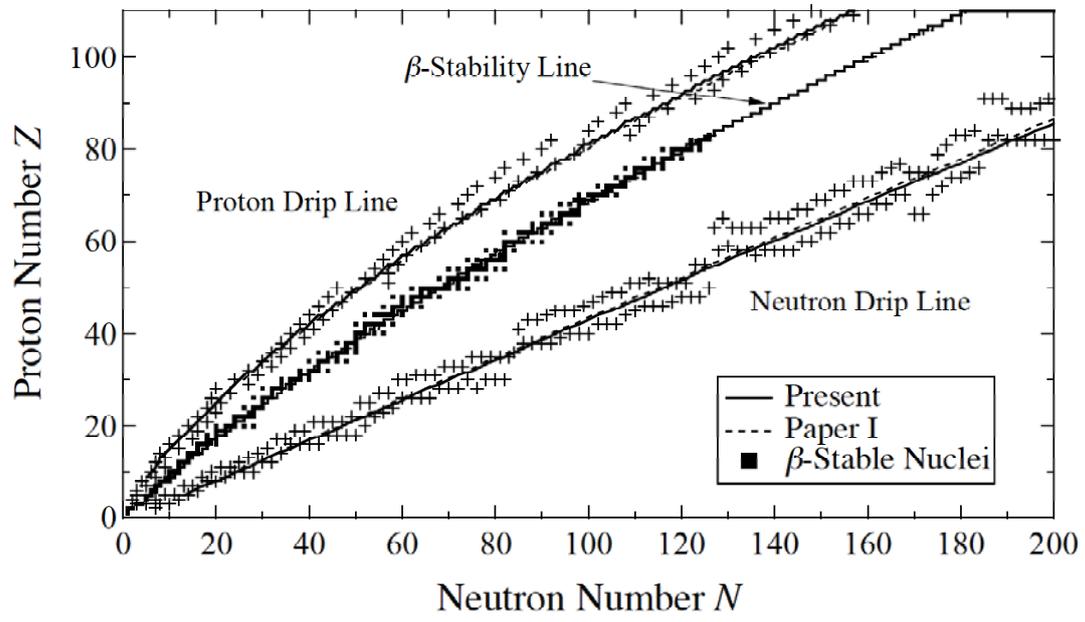

Fig. 4

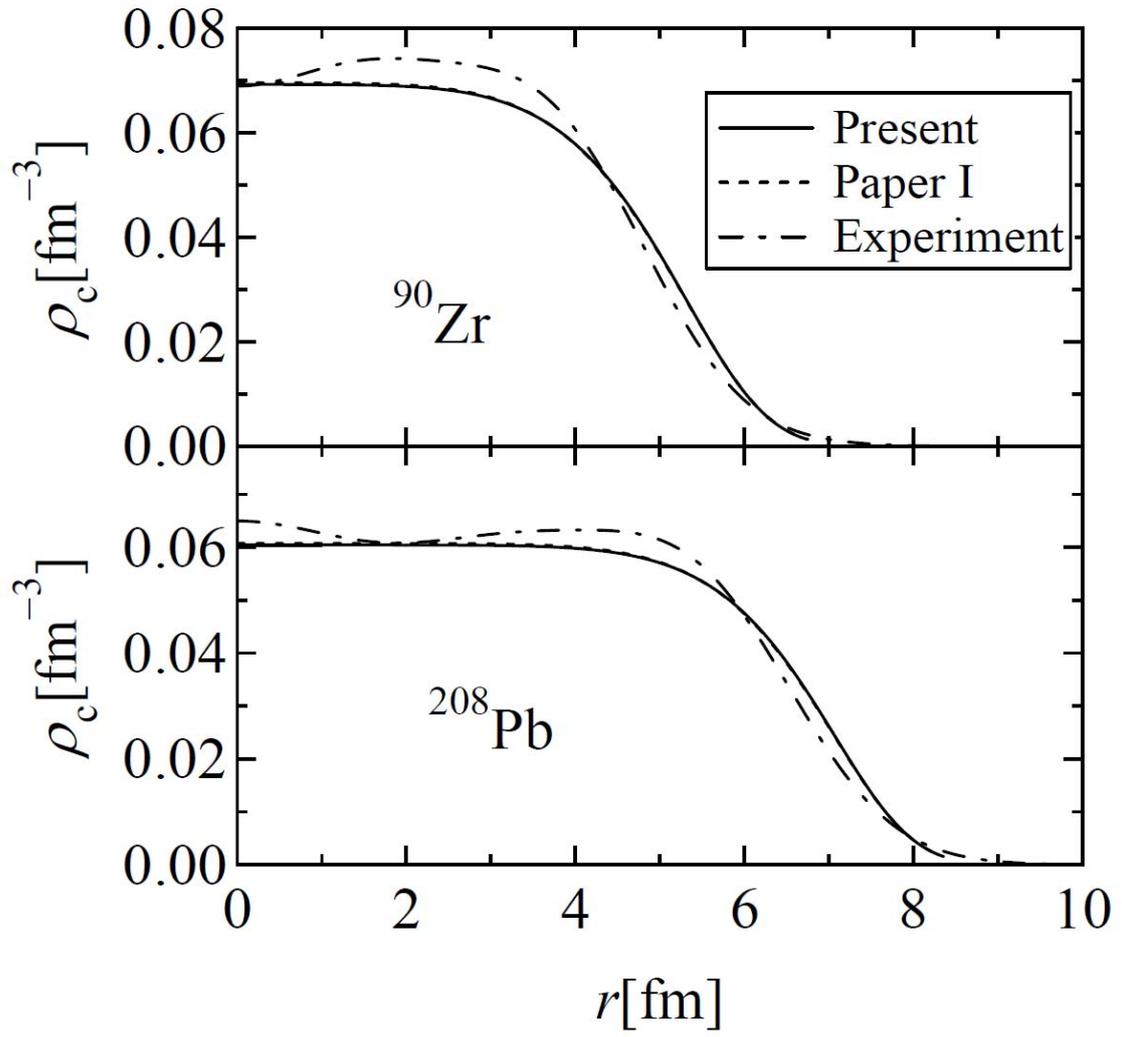

Fig. 5

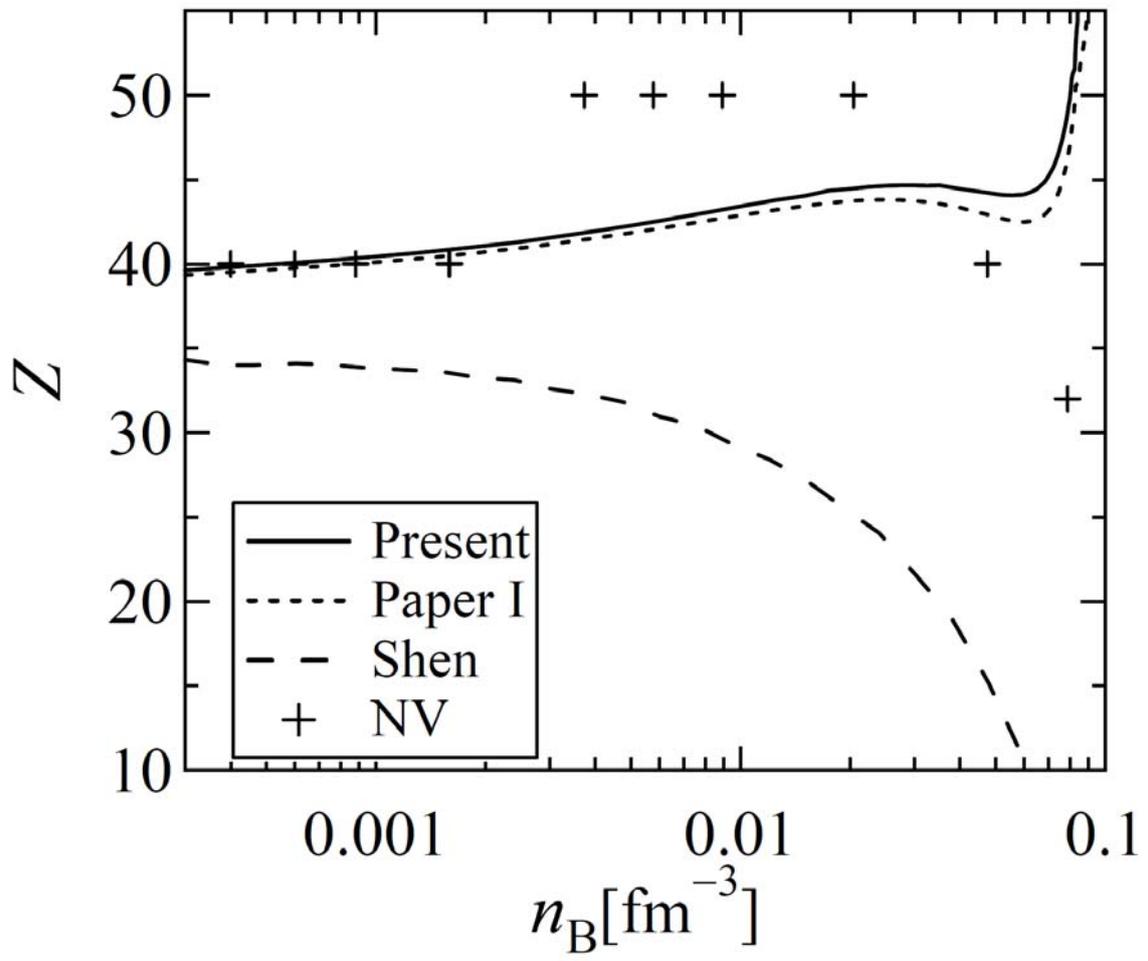

Fig.6

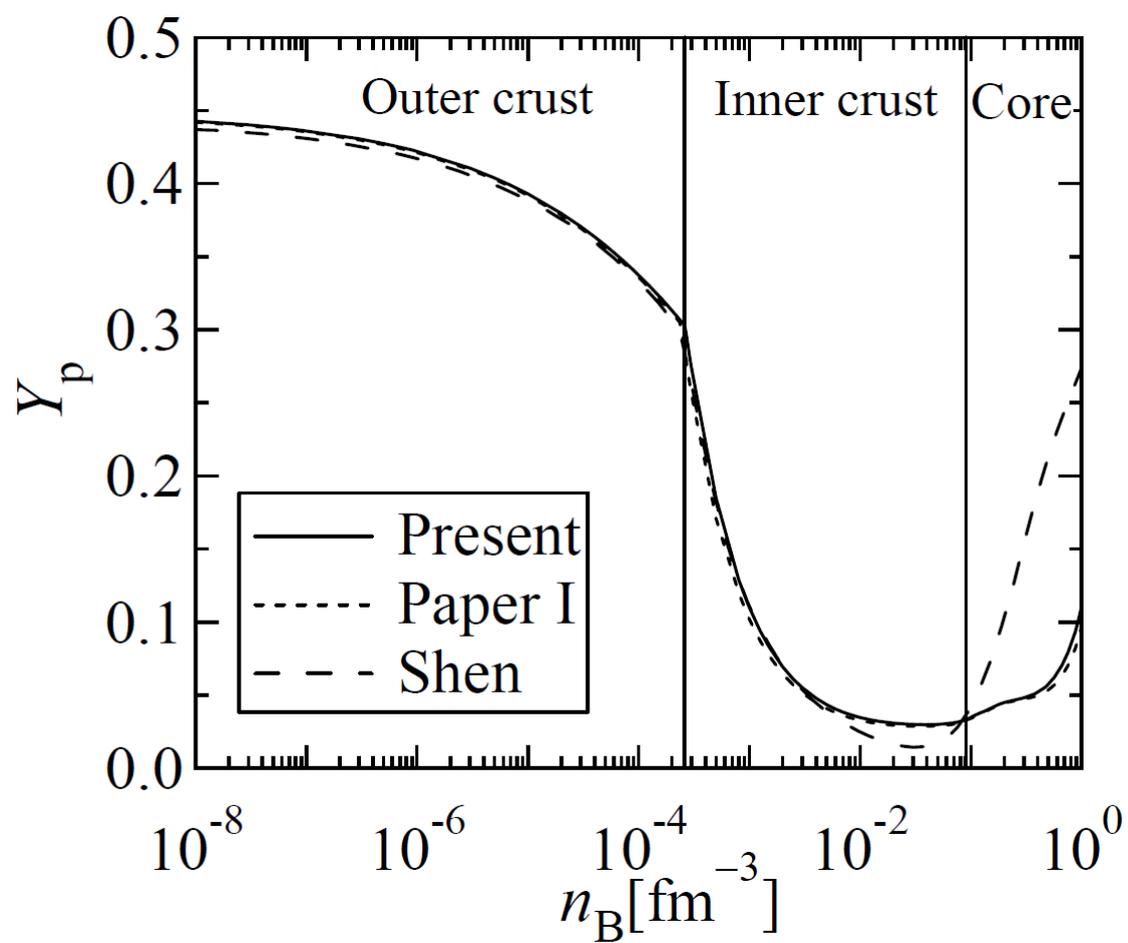

Fig7

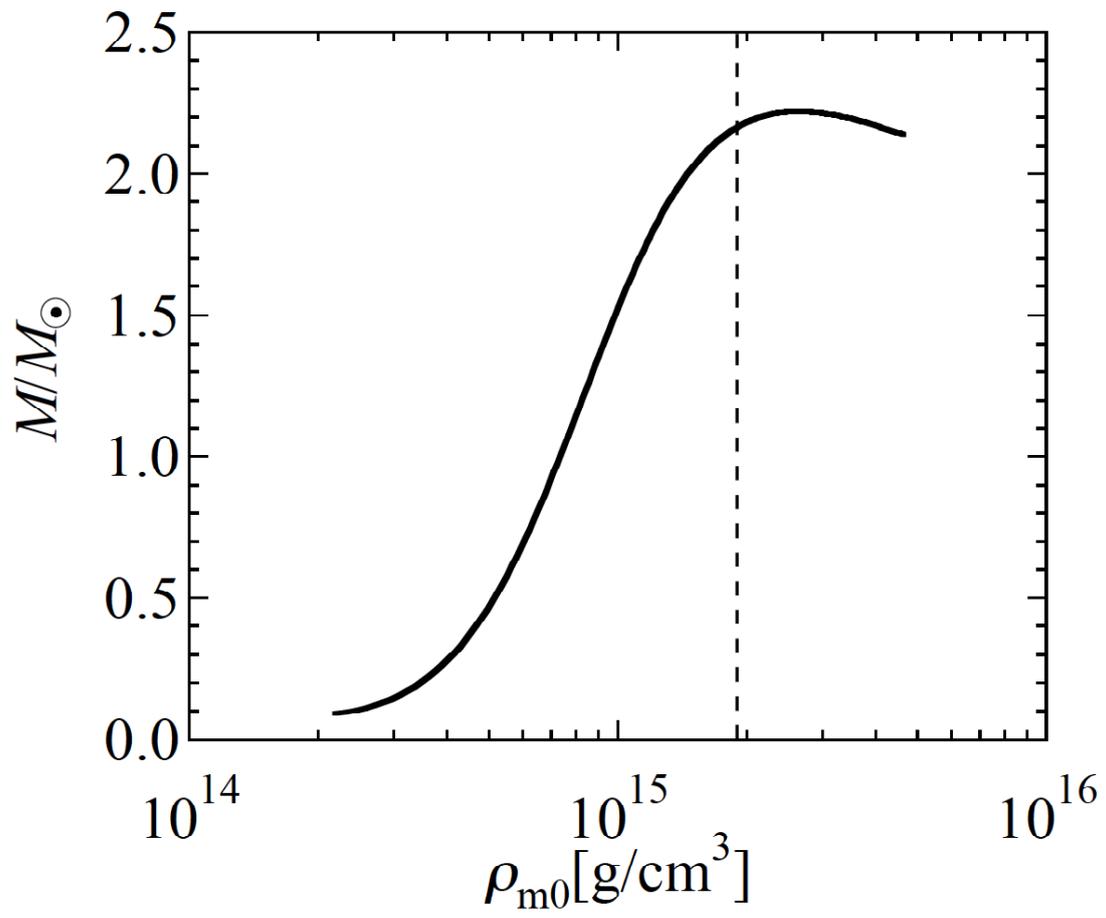

Fig.8

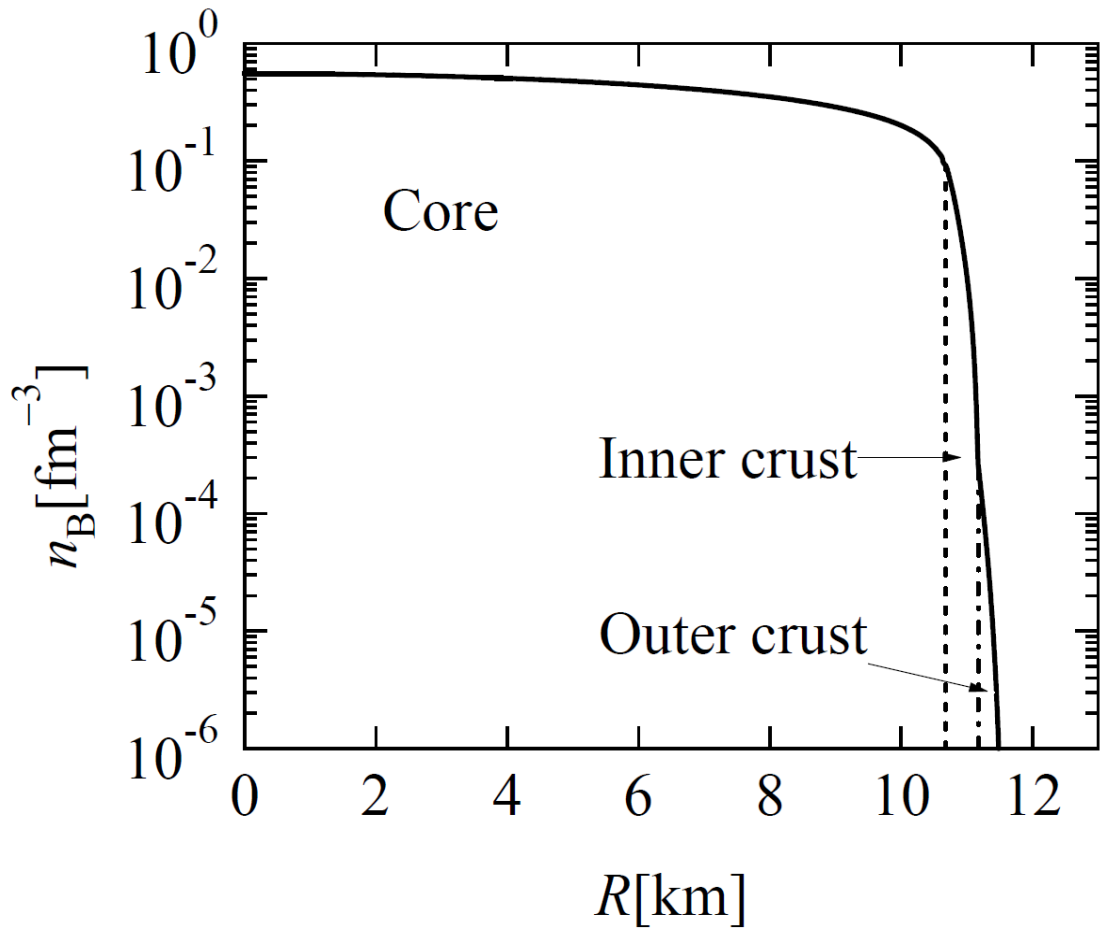

Fig.9